\documentclass[%
preprint,
 amsmath,amssymb,
 prl,
]{revtex4}

\usepackage{graphicx}% Include figure files
\usepackage{dcolumn}% Align table columns on decimal point

\usepackage[bf]{subfigure}
\usepackage{epsfig}

\usepackage{bm}% bold math
\begin{document}

%\preprint{Physics Review Letters}

\title{Circular Sound Wave Scattering Derivation for Acoustic Cloak Detection}% Force line breaks with \\
%\thanks{A footnote to the article title}%
%detection of acoustic cloak with a circular array 

 \author{Siyang Zhong} 
%  \altaffiliation[Also at ]{School of Engineering, University of Southampton, United Kingdom.}%Lines break automatically or can be forced with \\
%\author{Igor Vinogradov}%
% \email{Second.Author@institution.edu}
\author{Xun Huang}
 \email{huangxun@pku.edu.cn}
\affiliation{%
 State Key Laboratory of Turbulence and Complex Systems, College of Engineering, Peking University, Beijing, 100871, China
}
%\affiliation{
%2. School of Engineering, University of Southampton, SO 17 1BJ, United Kingdom
%}
%

%\collaboration{MUSO Collaboration}%\noaffiliation

%\author{Charlie Author}
 %\homepage{http://www.Second.institution.edu/~Charlie.Author}
%\affiliation{
 %Second institution and/or address\\
 %This line break forced% with \\
%}%
%\affiliation{
% Third institution, the second for Charlie Author
%}%
%\author{Delta Author}
%\affiliation{%
% Authors' institution and/or address\\
% This line break forced with \textbackslash\textbackslash
%}%

%\collaboration{CLEO Collaboration}%\noaffiliation

\date{\today}% It is always \today, today,
             %  but any date may be explicitly specified

\begin{abstract}
%

%
%The calculation of azimuthal sound localization resolution agrees well with the previous behavior %experimental results. 
In this Letter we develop analytical formulations to describe sound scattering in lossless medium due to 2D circular wave incident on an acoustic cloak. A perfect acoustic cloak is reflectionless and can completely hide the cloaked object from any sound waves. However, the realization of a perfect acoustic cloak is difficult. Compared to plane wave, our analytic calculations show that  circular wave from an annular line source generates distinct scattering patterns from an imperfect cloak design. Large modification in reflection directivities can be observed if the focal point of the incident wavefront is slightly customized. Hence, our work might find applications in acoustic cloak detection, which should have significant impact on cloak design and defense.

PACS number:  41.20.Jb, 42.25.Fx

\end{abstract}

% PACS, the Physics and Astronomy
%\pacs{}
                             % Classification Scheme.
%\keywords{Suggested keywords}%Use showkeys class option if keyword
                              %display desired
\maketitle

%\tableofcontents

A cloak bends wave fields in desired directions to shield any interior object from detection \cite{Cummer:08PRL, Chen:07Cloak, Cummer:07, Hetmaniuk:10Cloak}. The theory behind cloak design is  a conformal map that transforms a physical region with an interior cloaked hole to a virtual domain that is mathematically simply connected. 
This interesting principle is firstly proposed by Pendry \textit{et al.} \cite{Pendry:06Science} and Schurig \textit{et al.} \cite{Schurig:06Science} for electromagnetic wave cloaking. Cummer and Schurig presented a 2D acoustic cloaking design based on the isomorphism between acoustic equations and Maxwell’s equations \cite{Cummer:07}. 
Chen and Chan elegantly extended the transformation based acoustic cloak design to 3D spherical cloak cases \cite{Chen:07Cloak}. Cummer \textit{et al.} reported the same 3D design through a different route that relies on a  spherical harmonic scattering analysis \cite{Chen:07PRL}. Acoustic cloak designs are generally confirmed via numerical simulations with a plane, progressive, harmonic wave.  Some experimental demonstrations of acoustic cloak have been conducted for linear surface waves (at 10 Hz)  \cite{Farhat:08Cloak}, audible sound waves (at 1 to 3 kHz)  \cite{Popa:11Cloak}, and ultrasound waves (at 52 to 60 kHz) \cite{Zhang:11Cloak}.

In this Letter, we develop an analytical solution that describes the interaction between an acoustic cloak and  circular waves radiating from an annular line source.  Our results demonstrate the existence of peculiar scattering patterns if metamaterial properties \cite{Park:11PRL} of an acoustic cloak slightly differ from idealized ones. In addition, distinctively different scattering patterns can be found from  circular cloak shell simply by slightly adjusting the focal point of the wavefront. 
In contrast to normally incident plane wave, this annular line source setup might find applications in cloak detection.   

%No simplifications or assumptions are made except time harmonic dependence. 

Our investigation is conducted in an analytical way that can clearly provide physical insights.  Before starting the analytical derivations, we have need to recall the theoretical foundation behind acoustic cloak.  The propagation of linear sound wave perturbations is governed by linear Euler equations. With an $e^{-i\omega t}$ time dependence, these equations have the following form for a cloaked region,   
\begin{equation} \label{e:p}
i\omega p=\kappa\kappa_0\nabla\cdot\vec{v},
\end{equation}
\begin{equation}  \label{e:v}
\nabla p=i\omega\rho\rho_0\vec{v},
\end{equation}
\noindent where $p$ is sound pressure; $\vec{v}$ is the associated particle velocity;  $\kappa$ is anisotropic fluid bulk modulus relative to $\kappa_0$;  $\rho$ is anisotropic cloak density relative to fluid density $\rho_0$; $\rho_0$ and $\kappa_0$ are normalized to unity. For simplicity, the discussion in the following is focused on  circular cloaked shell  in 2D polar ($r-\theta$) coordinates (as shown in Fig. 1). The outer shell radius is $a$ and inner radius is $b$, which is nil in the transformed virtual region for a perfect 2D acoustic cloak. In other words, the transformation 
for an ideal cloak maps the physical region ($b<r<a$) to the virtual region ($0<r<a$). In the physical region, the cloak density is a second order tenser, that is, $\rho=\mathrm{diag}(\rho_r,\rho_{\theta})$. The relative anisotropic density and bulk modulus are \cite{Cummer:07}  
\begin{equation} \label{e:perfect}
\rho_r=\frac{r}{r-b}, \rho_{\theta}=\frac{r-b}{r}, \kappa=\left(\frac{a}{a-b}\right)^2\frac{r-b}{r}. 
\end{equation}
\noindent  Outside a cloak, $\rho_r$, $\rho_{\theta}$ and $\kappa$ are unity. %normalized
It is easy to see that the above design is impractical as $\rho_r$ will go infinity as $r$ approaches $b$. To avoid this potential singularity,  the physical region ($b<r<a$) can be mapped to a virtual region ($r_0<r<a$, $r_0>0$) by a linear transformation $f$. Accordingly, Norris  \cite{Norris:08Cloak} recently developed generic material specifications,  
\begin{equation}\label{e:gen}
\kappa(r)=\frac{1}{f^\prime}(\frac{r}{f})^{d-1},\rho_r=(\frac{r}{f})^{d-1}f^\prime, \rho_\theta=(\frac{r}{f})^{d-1}\frac{f^2}{r^2f^\prime},
\end{equation}
where $f^\prime=\mathrm{d}f/\mathrm dr$, and $d=2$ for 2D cases; $d=3$ for 3D cases.

%The time dependence of the pressure and the velocity can be taken into account simply by multiplying $e^{j\omega}t$. 

%and $\rho_0$ and $\kappa_0$ are unity and are omitted in this equation for brevity

%Figure~~??  shows three regions: (i) cloak .... 

%Of particular importance are harmonic waves.  

From Eqs.~(\ref{e:p})-(\ref{e:v}) it can be seen that harmonic acoustic pressure ($p$) inside acoustic cloak is governed by the following wave equation  
\begin{equation} \label{e:wave}
\nabla\cdot(\rho^{-1}\nabla p)+\frac{k^2}{\kappa}p=0,
\end{equation}
\noindent where $k$ is the normalized wavenumber.  In 2D polar coordinates, this wave equation has the following form,
\begin{equation} \label{e:sph}
 \frac{\partial}{\partial}\left(\frac{1}{\rho_r}\frac{\partial p}{\partial r}\right)+\frac{1}{r\rho_r}\frac{\partial p}{\partial r}+\frac{1}{r^2\rho_\theta}\frac{\partial^2p}{\partial\theta^2}-\frac{k^2}{\kappa}p=0. 
\end{equation}
\noindent Adopting the method of separation of variables, we let $p(r,\theta)=R(r)\Theta(\theta)$ and set $\Theta(\theta)$ as harmonic function, $\Theta(\theta)=e^{im\theta}$, where $m$ is an integer. Hence, $R(r)$ satisfies the following ordinary differential equation, 
\begin{equation} \label{e:bessel}
r\rho_\theta\frac{\partial}{\partial r}\left(\frac{r}{\rho_r}\frac{\partial R(r)}{\partial r}\right)+\left(\frac{k^2\rho_\theta r^2}{\kappa}-m^2\right)R(r)=0.
\end{equation}
\noindent From Eq.~(\ref{e:gen}) we have  $\rho_r=rf'/f$ and $\rho_\theta=f/(rf')$ for 2D cases, Eq.~(\ref{e:bessel}) becomes
\begin{equation} \label{e:bessel2}
\frac{\partial^2R(r)}{\partial r^2}+\frac{1}{f}\frac{\partial R(r)}{\partial f}+\left(k^2-\frac{m^2}{f^2}\right)R(r)=0.
\end{equation}
\noindent It is thus easy to find that the solution of Eq.~(\ref{e:bessel2}) has the form: $R(r)=B_mJ_m(kf(r))+C_mH_m(kf(r))$, where $J_m$ is the $m$th order Bessel function of the first kind, $H_m$ is the $m$th order Bessel function of the third kind, and $B_m$ and $C_m$ are parameters to be determined.  As a consequence, the sound pressure in an acoustic cloak is 
\begin{equation}\label{e:clk}
p^{clk}(r)=\sum_{m=-\infty}^{\infty}[B_mJ_m(kf(r))+C_mH_m(kf(r))]e^{im\theta}.
\end{equation}

Taking into account Sommerfeld radiation condition, the series forms of sound solutions in other regions are   
\begin{equation}
\begin{split}\label{e:wavs}
&p^{inc}=\sum_{m=-\infty}^{\infty}K_mJ_m(kr)e^{im\theta},\\
&p^{scat}=\sum_{m=-\infty}^{\infty}A_mH_m(kr)e^{im\theta},\\
%&p_{clk}=\sum_{m=-\infty}^{\infty}[B_mJ_m(kf(r))+C_mH_m(kf(r))]e^{im\theta},\\
&p^{int}=\sum_{m=-\infty}^{\infty}D_mJ_m(kr)e^{im\theta},
\end{split}
\end{equation}
\noindent where $p^{inc}$ is incident sound pressure, $p^{scat}$ is sound pressure scattered from an acoustic cloak, and $p^{int}$ is sound pressure in the interior of the cloaked hollow region.

The normal velocities can be derived using Eq.~(\ref{e:v}), which yields
\begin{equation} \label{e:normalV}
\begin{split}
&v^{inc}_r=\frac{k}{\rho_0}\sum_{m=-\infty}^{\infty}K_mJ^\prime_m(kr)e^{im\theta},\\
&v^{scat}_r=\frac{k}{\rho_0}\sum_{m=-\infty}^{\infty}A_m(H^\prime_m(kr))e^{im\theta},\\
&v^{clk}_r=\frac{kf'(r)}{\rho_0\rho_r}\sum_{m=-\infty}^{\infty}(B_mJ^\prime_m(kf(r))+C_mH^\prime_m(kf(r)))e^{im\theta},\\
&v^{int}_r=\frac{k}{\rho_0}\sum_{m=-\infty}^{\infty}D_mJ^\prime_m(kr)e^{im\theta},
\end{split}
\end{equation}
% 
%where the prime denotes differentiation with respect to the entire argument of the Bessel functions. 
\noindent where $^\prime$ stands for $\mathrm{d}/{\mathrm{d}r}$ and ${f'(r)}/{\rho_r}=({f}/{r})$ for 2D cases. 
Sound pressure and normal velocity should be continuous at cloak interfaces, that is, $r=a$ and $r=b$ in the physical region. We can therefore have the following relations, 
\begin{equation}
 \begin{split}
  &K_mJ_m(kb)+A_mH_m(kb)=B_mJ_m(kb)+C_mH_m(kb),\\
  &K_mJ'_m(kb)+A_mH'_m(kb)=B_mJ'_m(kb)+C_mH'_m(kb),\\
% \end{split}
%\end{equation}
%\begin{equation}
 %\begin{split}
  &B_mJ_m(kr_0)+C_mH_m(kr_0)=D_mJ_m(ka),\\
  &\frac{r_0}{a}(B_mJ'_m(kr_0)+C_mH'_m(kr_0))=D_mJ'_m(ka),
 \end{split}
\end{equation}
\noindent which yield 
\begin{equation}\label{e:sols}
\begin{split}
&A_m=\frac{K_m}{G}, B_m=K_m, C_m=\frac{K_m}{G},\\
&D_m=\frac{E_1E_6-E_3E_4}{E_2E_6-E_3E_5}K_m, 
\end{split}
\end{equation}
\noindent where the following formulas are used, 
\begin{equation}
 \begin{split}\label{e:rel}
 &E_1=J_m(kr_0), E_2=H_m(kr_0), E_3=J_m(ka),\\
 &E_4=\frac{r_0}{a}J'_m(kr_0), E_2=\frac{r_0}{a}H'_m(kr_0), E_3=J'_m(ka),\\
 &G=\frac{E_2E_6-E_3E_5}{E_1E_5-E_2E_4}. 
 \end{split}
\end{equation}

A 2D plane, progressive, harmonic wave can be described by $p^{inc}(x)=e^{ikx}$, which equals $e^{ikr \mathrm{cos} \theta}$ , that is, $\sum_{m=-\infty}^{\infty}i^{m}J_m(kr)e^{im\theta}$. Hence, the corresponding $K_m$ in Eq.~(\ref{e:wavs}) is $i^m$ for plane wave case.  Figure 1 shows a plane wave sound pressure field calculated from above series solutions. The 2D plane wave is incident from the left onto an acoustic cloak with $a=1$ and $b=0.5$. The normalized wavenumber is 10. For a perfect cloaking design, where $f: (b<r<a) \to (0<r<a)$, no reflection can be observed in Fig. 1(a). Since a perfect cloak design is difficult to implement, $f:  b \mapsto r_0, r_0>0$ is adopted in practical implementations.  For example, $r_0=0.02$ is represented in Fig. 1(b) by the dashed circle. However, scattering due to imperfect cloak design is now clearly visible in Fig. 1(b).

It is of interest to study the scattering patterns if a  circular sound wave is incident on an imperfect acoustic cloak. Figure 2 shows the setup of the problem.  The  circular wave incident on a cloak is generated by an annular line sound source, which presumably consists of numerous point sources obeying a uniform distribution. The radius of the annular line source is $R$, which is set to 3 in the following demonstration.  
The distance between the origin of the cloaked hollow region and the expected focal point of wavefront is $L$.  
The 2D annular line source gives 
\begin{equation}
J_0(kR)=\sum_{m=-\infty}^{\infty}J_m(kL)J_m(kr)e^{im\theta}, 
\end{equation}
\noindent that is,  $K_m=J_m(kL)$ for the  circular wave case. Without the presence of an cloak, the origin of the line source will be the focal point of wavefronts. In the following demonstrations, we slightly move the focal point along the focal path in Fig.~\ref{f:Setup}. Practical implementation can be realized by marrying the concept of beamforming (in array signal processing) \cite{Xun:11JASA, Xun:12JASA} with harmonic analysis introduced in this Letter. 

In our demonstration, Eqs.~(\ref{e:clk})-(\ref{e:rel}) can be used to calculate sound propagation and reflection.  Figure~\ref{f:spheric} shows sound reflections due to  circular waves with $L=0.1$ and $L=0.3$, respectively. It can be seen that sound reflections are symmetric with respect to the $x$-axis. The resultant scattering patterns are quite different than plane wave cases. In addition, distinctive scattering patterns appear if the focal point of  circular wavefront is modified. When $L=0.1$, the directivity of sound reflections are approximately in the $x$ direction. When $L=0.3$, sound reflections move around the $y$ direction.  For imperfect cloak cases in Fig. 1(b),  it can be seen that almost no sound energy reaches the interior region. In contrast, it appears that a small fraction of sound energy arrives the cloaked hollow region. More details of sound transmission and reflections can be found in the online supplementary materials (including animations of plane wave and  circular wave cases).  It is for this reason that our work might find applications in imperfect acoustic cloak detection.

In summary, the method proposed in this Letter permits a purely analytical design of acoustic cloak detection strategy.
Using 2D harmonic analysis, our study demonstrates unique scattering pattern due to incident  circular waves generated by an annular line source. We have also developed 3D formulations, which are omitted here for brevity. In this work we conducted 2D simulations using the proposed analytical formulations at various $r_0$ and $L$ and similar conclusions can be drawn.  It is worthwhile to note that the series solution works to any wavelength condition. Hence, we conclude that an annular line source might be useful in detecting \cite{Zhang:10PRL} an enclosed imperfect acoustic cloak, which should have a significant impact on acoustic cloak design and defense.

%It is also important to note that 

%A strategy detecting an acoustic cloak should have significant impact on  both practical needs and academic interests. 
%A few investigations have been conducted for electromagnetic cloak.

The preparation of this Letter is partially supported by the NSF Grant of China (Grant Nos. 11172007 and 11110072) and SRF for ROCS, SEM.

%\section*{APPENDIX: 3D FORMULATIONS}

%
\bibliographystyle{apsrev4-1}
\bibliography{CloakPRL}

\newpage

\begin{figure}[htp]
\begin{center}
\begin{minipage}[t]{165mm}
    \subfigure[\label{f:monopolea}]
    {\includegraphics[clip,width=80mm]{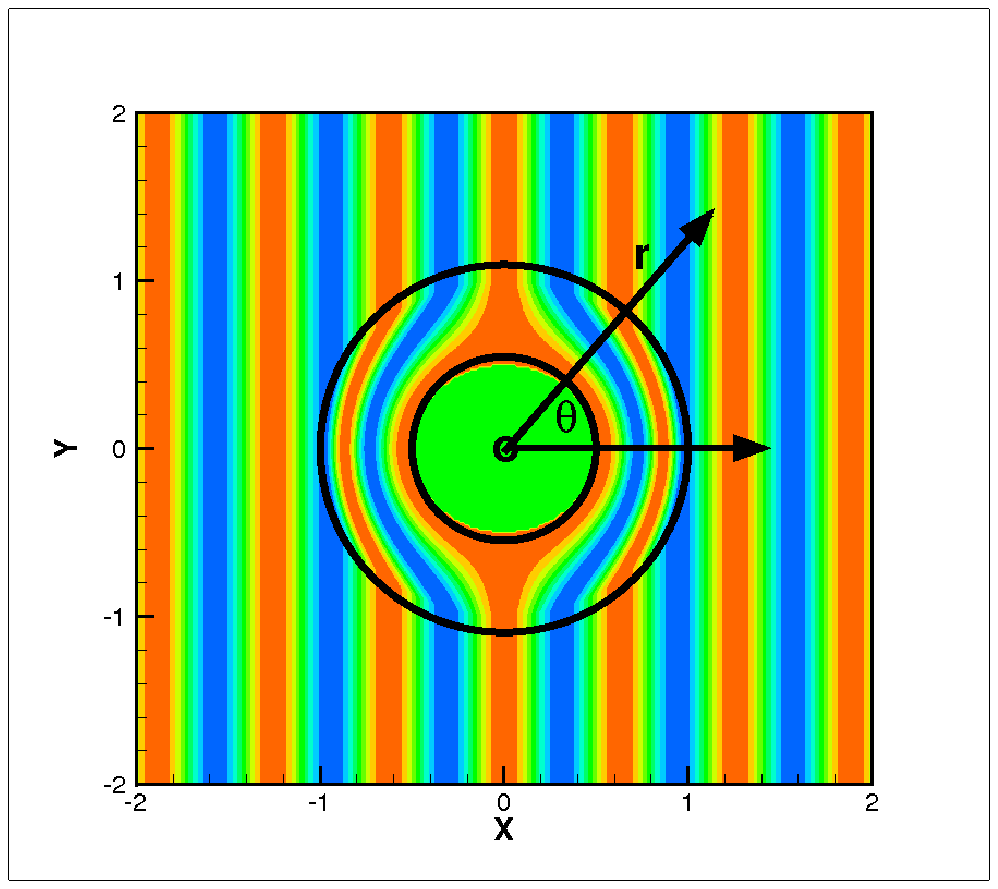}}
    \subfigure[\label{f:monopoleb}]
    {\includegraphics[clip,width=80mm]{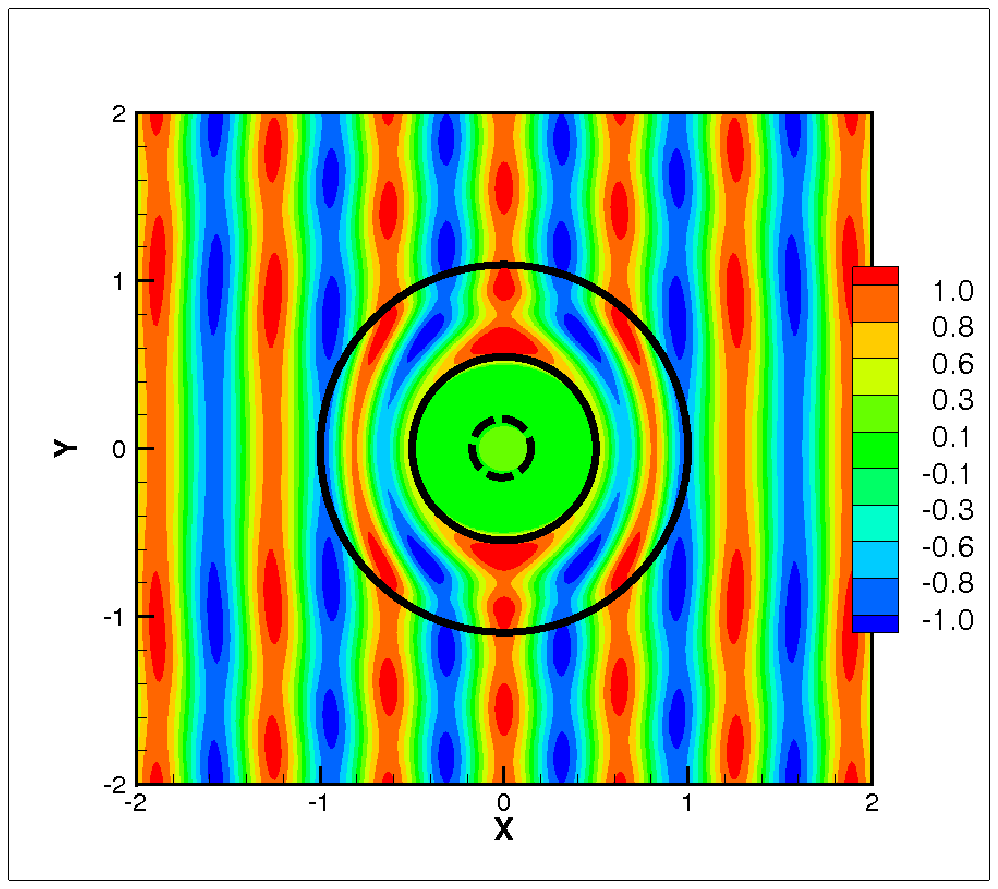}}
    \caption{ The real part of sound pressure field in the physical region. The plane wave is incident from the left. (a) Perfect cloak case, material properties by Eq.~(\ref{e:perfect}). (b) Imperfect cloak, material properties by Eq.~(\ref{e:gen}), $f$ maps $b=0.5$ to $r_0=0.02$. 
  }\label{f:1}
\end{minipage}
\end{center}
\end{figure} 

\begin{figure}[h!]
\begin{center}
    \includegraphics[width=80mm]{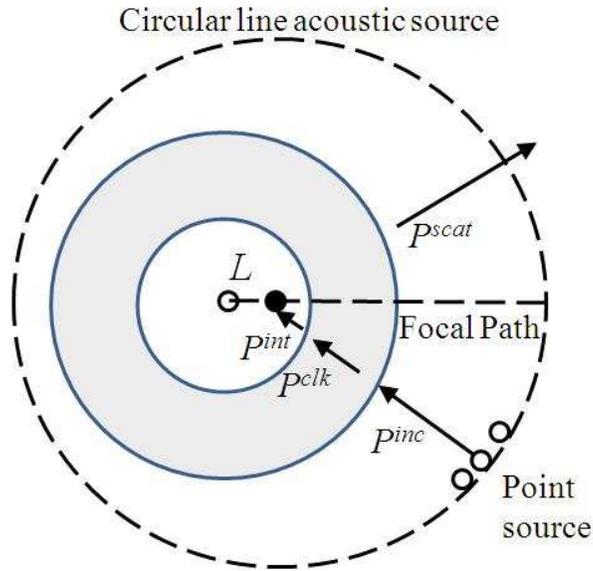}
    \caption {\label{f:Setup}  Configuration of  circular wave from an annular line source incident on an acoustic cloak.   }
\end{center}
\end{figure}

\begin{figure}[htp]
\begin{center}
\begin{minipage}[t]{165mm}
    \subfigure[\label{f:monopolea}]
    {\includegraphics[clip,width=80mm]{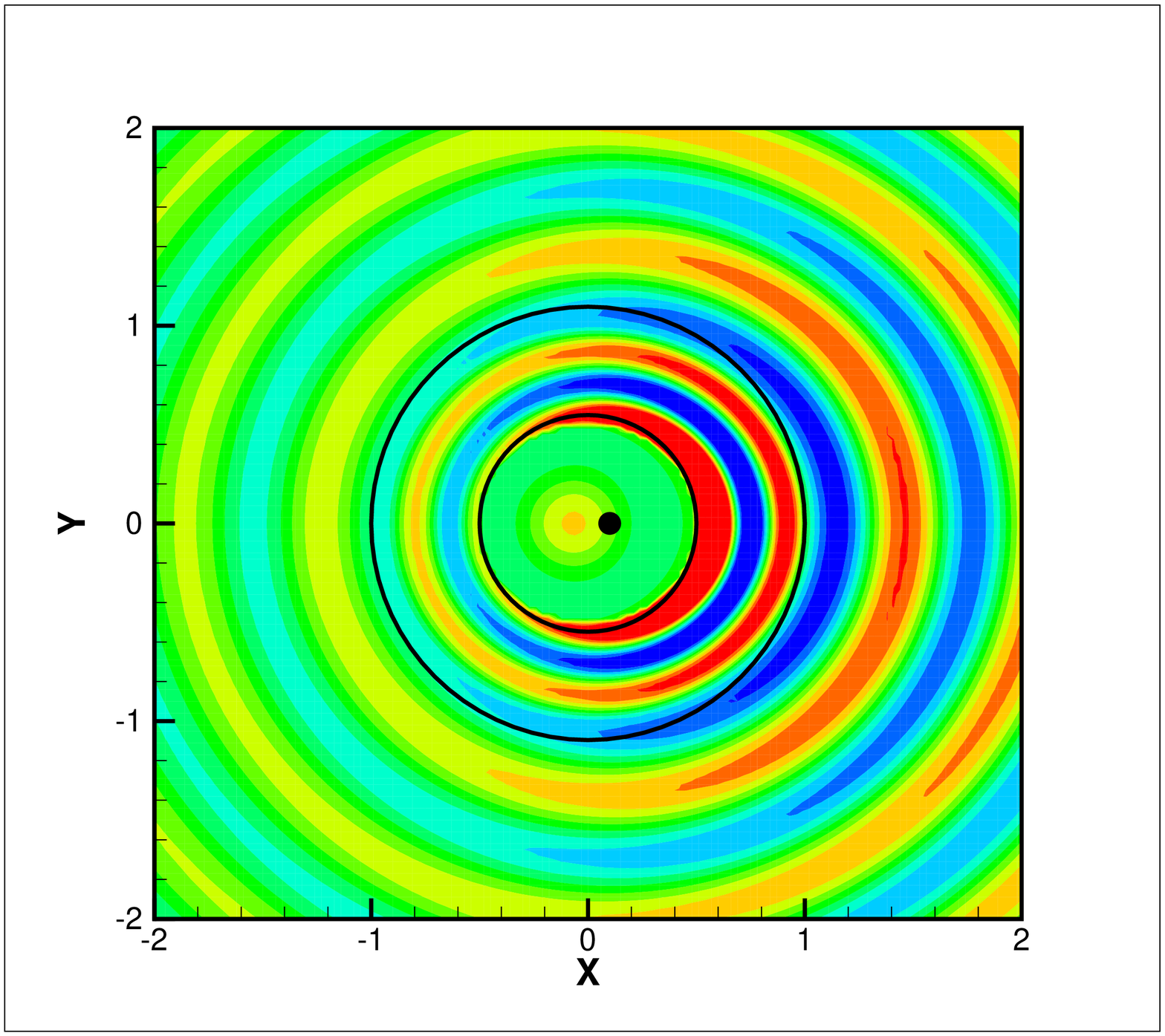}}
    \subfigure[\label{f:monopoleb}]
    {\includegraphics[clip,width=80mm]{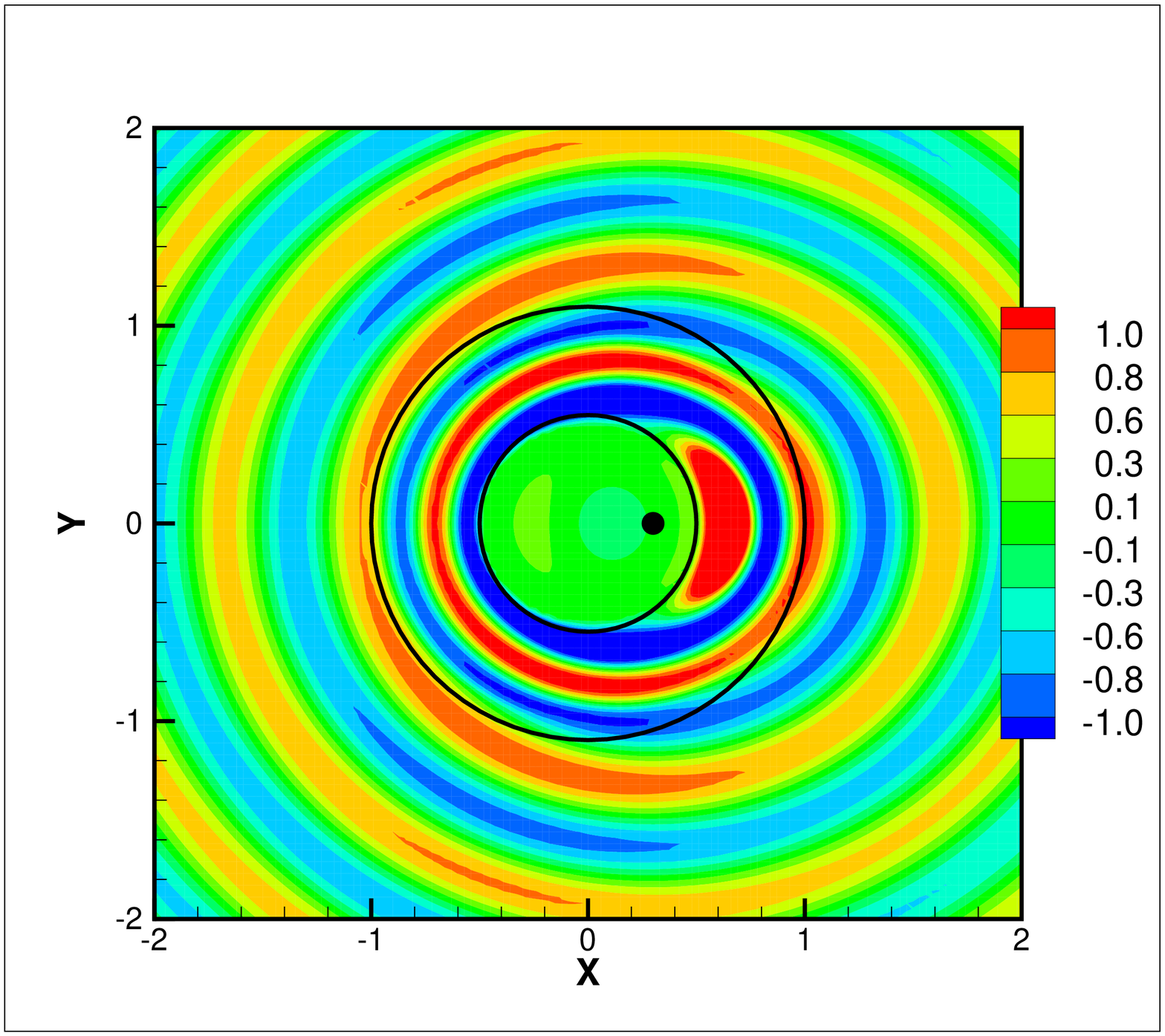}}
    \caption{ Sound scattering due to  circular waves incident on an imperfect acoustic cloak  by Eq.~(\ref{e:gen}), $f$ maps $b=0.5$ to $r_0=0.02$. (a) $L=0.1$. (b) $L=0.3$, where the black solid dot stands for the expected focal point.      }\label{f:spheric}
\end{minipage}
\end{center}
\end{figure} 

\iffalse
%

\begin{figure}[h!]
\begin{center}
    \includegraphics[width=140mm]{figures/Setup2.eps}
    \caption {\label{f:Setup} The sketch of the numerical setups  (not to scale), where five $\circ$ on each cochlea represent 300 sound sensors in the simulations.     }
\end{center}
\end{figure}

\begin{figure}[h!]
\begin{center}
    \includegraphics[width=140mm]{figures/ circular.eps}
    \caption {\label{f: circular} The simulated sound localization results, where (a)(b) $f=2\,$kHz, (c)(d) $f=5\,$kHz, with (a)(c) straightened cochleae and (b)(d) spiral cochleae.        }
\end{center}
\end{figure}

\begin{figure}[h!]
\begin{center}
    \includegraphics[width=160mm]{figures/ViewAngle.eps}
    \caption {\label{f:ViewAngle} Simulated beam patters with the straightened and the spiral cochleae, where (a) the vertical profiles and (b) the horizontal profiles.        } \label{f:}
\end{center}
\end{figure}

\fi
\end{document}